\documentclass[aps,prl,twocolumn,showpacs,superscriptaddress]{revtex4}
\usepackage{amssymb}
\usepackage{amsfonts}
\usepackage{graphicx}
\usepackage{CJK}
\usepackage{indentfirst}
\usepackage{amsmath}
\usepackage{epstopdf}
\usepackage{array}
\begin{document}


\title{Observation of Two-dimensional Superconductivity in an Ultrathin Iron-Arsenic Superconductor}

\author{Chi Zhang}
\affiliation{State Key Laboratory of Functional Materials for
Informatics, Shanghai Institute of Microsystem and Information
Technology, Chinese Academy of Sciences, Shanghai 200050,
China}\affiliation{CAS Center for Excellence in Superconducting
Electronics(CENSE), Shanghai 200050, China}\affiliation{University
of Chinese Academy of Sciences, Beijing 100049, China}

\author{Tao Hu}\email[]{hutao@baqis.ac.cn}
\affiliation{Beijing Academy of Quantum Information Sciences, Beijing 100193, China}

\author{Teng Wang} \affiliation{State Key
Laboratory of Functional Materials for Informatics, Shanghai
Institute of Microsystem and Information Technology, Chinese Academy
of Sciences, Shanghai 200050, China}\affiliation{CAS Center for Excellence in Superconducting
Electronics(CENSE), Shanghai 200050, China}\affiliation{School of Physical Science and Technology, ShanghaiTech University, Shanghai 201210, China}

\author{Yufeng Wu}
\affiliation{Beijing Academy of Quantum Information Sciences, Beijing 100193, China}

\author{Aobo Yu}
\affiliation{State Key Laboratory of Functional Materials for
Informatics, Shanghai Institute of Microsystem and Information
Technology, Chinese Academy of Sciences, Shanghai 200050,
China}\affiliation{CAS Center for Excellence in Superconducting
Electronics(CENSE), Shanghai 200050, China}\affiliation{University
of Chinese Academy of Sciences, Beijing 100049, China}

\author{Jianan Chu}
\affiliation{State Key Laboratory of Functional Materials for
Informatics, Shanghai Institute of Microsystem and Information
Technology, Chinese Academy of Sciences, Shanghai 200050,
China}\affiliation{CAS Center for Excellence in Superconducting
Electronics(CENSE), Shanghai 200050, China}\affiliation{University
of Chinese Academy of Sciences, Beijing 100049, China}

\author{Han Zhang}
\affiliation{State Key Laboratory of Functional Materials for
Informatics, Shanghai Institute of Microsystem and Information
Technology, Chinese Academy of Sciences, Shanghai 200050,
China}\affiliation{CAS Center for Excellence in Superconducting
Electronics(CENSE), Shanghai 200050, China}\affiliation{University
of Chinese Academy of Sciences, Beijing 100049, China}

\author{Hong Xiao} \affiliation{Center for High Pressure Science and Technology Advanced Research, Beijing 100094, China}

\author{Wei Peng} \affiliation{State Key Laboratory of Functional Materials for
Informatics, Shanghai Institute of Microsystem and Information
Technology, Chinese Academy of Sciences, Shanghai 200050,
China}\affiliation{CAS Center for Excellence in Superconducting
Electronics(CENSE), Shanghai 200050, China}\affiliation{University
of Chinese Academy of Sciences, Beijing 100049, China}

\author{Zengfeng Di} \affiliation{State Key Laboratory of Functional Materials for
Informatics, Shanghai Institute of Microsystem and Information
Technology, Chinese Academy of Sciences, Shanghai 200050,
China}\affiliation{CAS Center for Excellence in Superconducting
Electronics(CENSE), Shanghai 200050, China}\affiliation{University
of Chinese Academy of Sciences, Beijing 100049, China}

\author{Shan Qiao}
\affiliation{State Key Laboratory of
Functional Materials for Informatics, Shanghai Institute of
Microsystem and Information Technology, Chinese Academy of Sciences,
Shanghai 200050, China}\affiliation{CAS Center for Excellence in Superconducting
Electronics(CENSE), Shanghai 200050, China}\affiliation{University of Chinese Academy of Sciences, Beijing 100049, China}

\author{Gang Mu}
\email[]{mugang@mail.sim.ac.cn} \affiliation{State Key Laboratory of
Functional Materials for Informatics, Shanghai Institute of
Microsystem and Information Technology, Chinese Academy of Sciences,
Shanghai 200050, China}\affiliation{CAS Center for Excellence in Superconducting
Electronics(CENSE), Shanghai 200050, China}\affiliation{University of Chinese Academy of Sciences, Beijing 100049, China}

\begin{abstract}
Two-dimensional (2D) superconductors supply important platforms for exploring new quantum physics and high-$T_c$ superconductivity.
The intrinsic superconducting properties in the 2D iron-arsenic superconductors are still unknown owing to the difficulties in the preparation of ultrathin samples.
Here we report the fabrication and physical investigations of the high quality single-crystalline ultrathin films of the iron-arsenic superconductor KCa$_2$Fe$_4$As$_4$F$_2$.
For the sample with the thickness of 2.6$\sim$5 nm (1$\sim$2 unit cells), a sharp superconducting transition at around 30 K (onset point) is observed.
Compare with the bulk material, the ultrathin sample reveals a relatively lower $T_c$, wider transition width, and broader flux liquid region under the in-plane field.
Moreover, the angle dependent upper critical field follows the Tinkham model, demonstrating the two-dimensional superconductivity in ultrathin KCa$_2$Fe$_4$As$_4$F$_2$.
\end{abstract}

 \maketitle

\section*{1 Introduction}
Two-dimensional (2D) superconductivity draw great interest due to the emergence of new quantum phenomena~\cite{Reyren1196,Gozar,Saito2016}, including Ising superconductivity~\cite{Ising,Ising-2,Ising-3},
quantum metallic state~\cite{Bose-metal1,Bose-metal2,Bose-metal3,Bose-metal4}, Berezinskii-Kosterlitz-Thouless (BKT)
transition~\cite{BKT-1,BKT-2}, and even the significant enhancement of $T_c$~\cite{FeSe,TaS2-1,TaS2-2}. From the material point of view, most of the 2D superconductors (2DSCs) under investigation belong to
the transition metal disulfide compounds~\cite{Ising,Ising-2,Ising-3,TaS2-1,TaS2-2},
which have the easy-to-exfoliated layered structure. Owing the unconventional superconducting (SC) properties and the high-$T_c$, the 2D behaviors of cuprates~\cite{Bednorz1986} and iron-based superconductors~\cite{LaFeAsO}
are very deserve to study.
For the cuprates, mechanical exfoliation method was adopted to create the ultrathin Bi$_2$Sr$_2$CaCu$_2$O$_{8+x}$ (Bi2212) superconductor down to half unit-cell in recent years, where a rather high-$T_c$ very close to the
bulk compound and the absence of dimensionality effect were revealed~\cite{Jiang2014,Yu2019}. As for the iron-based superconductors, investigations focus on the iron-selenic (FeSe, including FeSe$_{1-x}$Te$_x$) materials by the method of
the precisely controlled molecular beam-epitaxy (MBE)
growth~\cite{FeSe} or mechanical exfoliation~\cite{Tang_2019}. During this process, high temperature superconductivity was observed in the single-layer FeSe film grown on SrTiO$_3$ substrates~\cite{FeSe}.
On the other hand, the iron-arsenic system has a more diverse crystal structure and higher $T_c$~\cite{LaFeAsO,Sm55K,Zhu2009,122,12442-1}.
Thus, clarifying the performance in terms of dimensionality effect and $T_c$ enhancement in this iron-arsenic system is vary
crucial in understanding the intrinsic properties in the 2D limit.

Due to the stronger inter-layer coupling, as compared with the Bi2212 system, it has been a huge challenge for a long time to obtain the ultrathin samples in the iron-arsenic system.
The recently reported 12442 system~\cite{12442-1,12442-2,12442-3,12442-4,12442-5}, which has double FeAs layers between neighboring insulating layers,
reveals a very large resistivity anisotropy ratio ($\rho_c/\rho_{ab} \approx 3150$)~\cite{WangCrystal}. This value is much larger than those of other iron-arsenic superconductors~\cite{Zverev_2011,Xiao_2016},
indicating a rather weak inter-layer coupling. Using an
improved mechanical exfoliation method, we successfully fabricated the ultrathin films of KCa$_2$Fe$_4$As$_4$F$_2$ (K-12442) with the thickness down to 1$\sim$2 unit cells. Here we report the mechanical exfoliation and superconducting properties
of the ultrathin K-12442, which are compared with the bulk samples. By examining the angle dependence of the upper critical field, a clear two-dimensional feature is revealed in the ultrathin samples, which is in sharp contrast with
the bulk samples.
The present work provides an important platform to study the dimensionality effect of unconventional superconductors, which is rather different with the cuprate system.

\begin{figure}
\includegraphics[width=8.5cm]{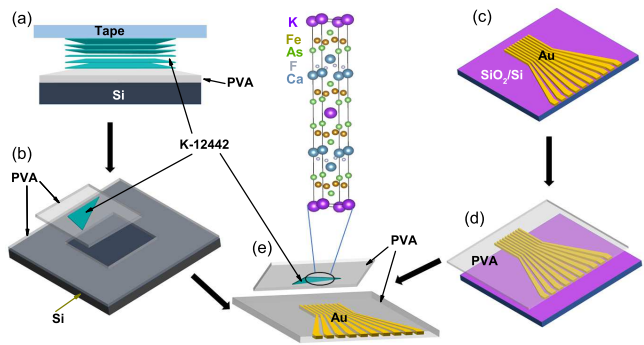}
\caption {(color online) The fabrication process for device of ultrathin K-12442. (a) Prefabricated Au electrodes on Si/SiO$_2$ substrate by ultraviolet lithography and metal thermal evaporation.
(b) peeling off the Au electrodes. (c) PVA-assisted mechanical exfoliation of K-12442 sample. (d) Transferring the PVA film and ultrathin K-12442 sample from the substrate.
(e) The upper and lower layers of PVA film encapsulate the sample and Au electrodes to obtain device of  ultrathin K-12442. } \label{fig1}
\end{figure}

\section*{2 Materials and methods}
The K-12442 single crystals were grown by the
self-flux method~\cite{WangCrystal-2}. We adopted an improved mechanical exfoliation method to fabricate the ultrathin samples of K-12442.
The traditional process of mechanical exfoliation uses the silicon-based (SiO$_2$/Si) substrate~\cite{Jiang2014,Yu2019}, which has insufficient adhesion to some materials with a strong interlayer coupling.
Thus it is difficult to obtain ultra-thin layers by this traditional method in many quasi-2D materials. Here we use an organic substrates, polyvinyl alcohol (PVA),
to replace silicon-based substrates to enhance the adhesion without changing the characteristics of the materials. PVA has a relatively strong adhesion and is typically used to transfer 2D materials~\cite{9052728}.
The advantage of strong adhesion for this material is utilized in the process of mechanical exfoliation in this work.
Figure 1 shows the fabricating process for ultrathin K-12442 sample devices. First, the self-made 8\% concentration of PVA solution is
dripped on a silicon substrate to cover the entire substrate surface, which stands for 5 hours in a drying oven to evaporate the water in the PVA solution naturally.
In this way we obtained a smooth PVA film with a certain thickness and adhesion on the silicon substrate. Next, a fresh surface of layered K-12442 crystal was cleaved from blue tape and put it onto the PVA substrate.
Afterwards, a more adhesive tape was used to thin the K-12442 flakes on the PVA substrate for 2-3 times until we obtained the ultrathin K-12442 sample we needed. For the prefabricated Au electrodes, we use ultraviolet
lithography and metal thermal evaporation technology to produce Au electrode patterns on silicon wafer (with 300 nm SiO$_2$). Similar to the above process, the PVA solution is dripped on the silicon wafer
with the prefabricated Au electrodes to obtain a PVA film with the thickness of several hundred micrometers.The ultrathin K-12442 sample and the prefabricated Au electrodes can be transferred by
the PVA film from the silicon and SiO$_2$/Si substrates respectively. Using the three-axis alignment platform, the ultrathin sample and the
Au electrodes can be aligned, which are encapsulated between two PVA sheets. Because the ultrathin K-12442 samples are sensitive to the external environment, the packaging of the samples must be completed within half an
hour under the atmospheric condition. In the upper left inset of Fig. 2, we show a optical microscopic picture of the ultrathin sample denoted as S-2. One can see that the sample is transparent and the length of the sample can
be as large as 30 $\mu$m.

The electrical transport data were collected by the
standard four-probe method using the prefabricated Au electrodes (see Fig. 1). The external magnetic field was rotated in the plane
perpendicular to the electric current. $\theta$ denoted the included angle
between external field $H$ and $c$ axis of the crystal. The applied
electric current is 10 $\mu$A.

\begin{figure}
\includegraphics[width=9.5cm]{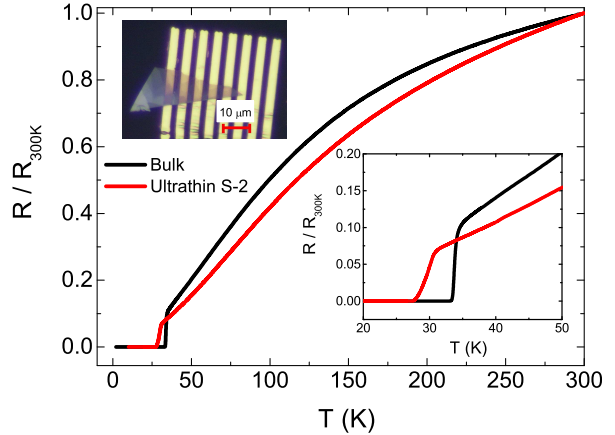}
\caption {(color online) Temperature dependence of normalized resistance of K-12442 ultrathin sample S-2. The data of the bulk sample is also shown for comparison. The upper left
inset shows the picture of the sample under optical microscope. The yellow strips are the strip electrodes. The lower right inset is an enlarged view of the resistance data.} \label{fig2}
\end{figure}

\begin{figure*}
\includegraphics[width=12cm]{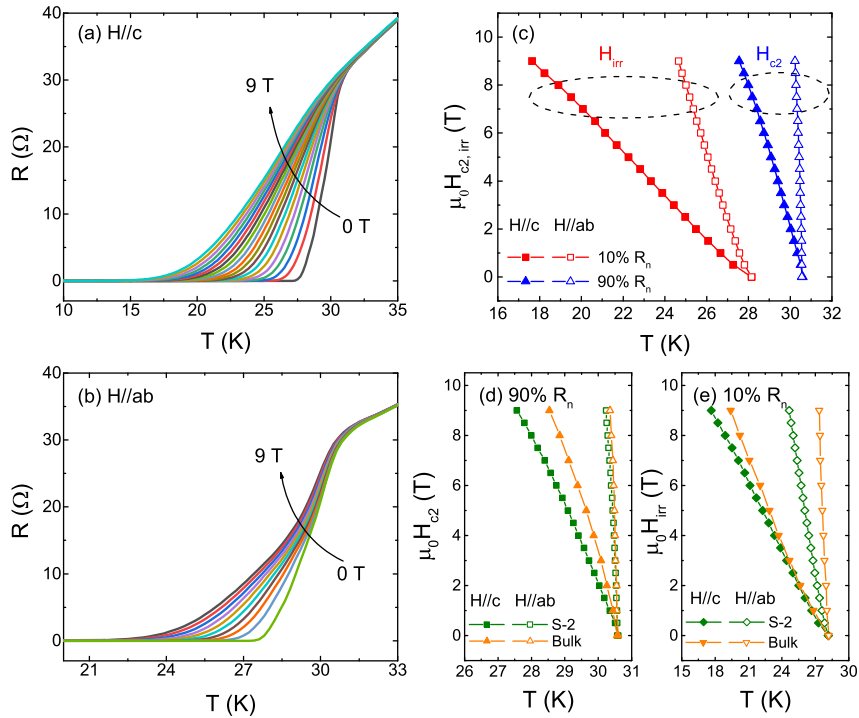}
\caption {(color online) (a, b) Electrical resistance of the ultrathin sample S-2 as a function of temperature under the magnetic field up to 9 T with $H$//c and $H$//ab, respectively.
The increasing steps of the field is 0.5 T for the two figures. (c) Upper critical fields $H_{c2}$ and irreversible field
$H_{irr}$ of the ultrathin sample as a function of temperature for two different orientations. (d, e) Comparison of $H_{c2}$ and $H_{irr}$ between the ultrathin and bulk samples. In order to have an intuitive comparison,
the data of the bulk sample are shifted to lower temperatures by 2.75 K and 4.50 K for (d) and (e) respectively.}
\label{fig3}
\end{figure*}

\section*{3 Results and discussion}
Normalized resistance of ultrathin K-12442 sample S-2, as well as the data of the bulk sample, is shown in Fig. 2. The thickness of this sample is estimated to be 2.6$\sim$5 nm, corresponding to 1$\sim$2 unit cells (see SI).
Typically the temperature dependent tendency of the ultrathin sample is
very similar to that of the bulk sample, both of which reveal a negative curvature with residual resistance ratio (RRR) above 10. The onset SC transition temperature $T_c^{onset}$ for the ultrathin sample is 30.6 K determined using a
criterion of 90\% $R_n$ ($R_n$ is the normal state resistance at the SC transition point). This value is slightly lower than that of the bulk sample. The more significant difference between the ultrathin and bulk samples is
the SC transition width $\Delta T_c$. Defined as the width between 10\%$R_n$ and 90\%$R_n$, the vary small $\Delta T_c = 0.6$ K in the bulk sample is increased to 2.4 K in the ultrathin sample.
Such a dimensionality effect on $T_c$ and  $\Delta T_c$ may reflect the enhancement of quantum fluctuation by the decrease of the sample thickness, which is in sharp contrast to that observed in ultrathin cuprates.
In the case of Bi2212, the transition width $\Delta T_c$ is as large as 10 K in both the bulk and monolayer samples. Such a comparison evidently indicates that the interlayer coupling in the iron-arsenic superconductors
is stronger than that of cuprates, which leads to a more prominent dimensionality effect in the former system.

\begin{table*}
\centering \caption{Summary of the detailed parameters about $T_c$ and upper critical fields for both the ultrathin and bulk K-12442.}
\medskip
\begin{tabular}{p{2cm}<{\centering}p{1.5cm}<{\centering}p{1.5cm}<{\centering}p{1.5cm}<{\centering}p{2.8cm}<{\centering}p{2.8cm}<{\centering}}
\hline
Sample & $T_c^{onset}$ & $T_c^{zero}$ & $\Delta T_c$   & $d\mu_0H_{c2}^{ab}(T)/dT|_{T_c}$ & $d\mu_0H_{c2}^{c}(T)/dT|_{T_c}$   \\
 & (K) & (K) & (K) & (T/K) &(T/K)\\
\hline
Ultrathin  & 30.6  & 28.2  & 2.4  & -47.6 & -3.3  \\
Bulk  &   33.3  & 32.7  & 0.6  & -50.9 & -6.4 \\
\hline
\end{tabular}
\end{table*}

We next focus on the critical fields $H_{c2}$ and irreversible field
$H_{irr}$ of the ultrathin sample at various field-orientations. As shown in Figs. 3(a) and (b), the $R-T$ curves becomes more broadening with the increase of magnetic field, which is more more significant when the field
is parallel to the $c$ axis. This is qualitatively consistent with that observed in bulk samples~\cite{WangCrystal-2,Wang2020}. We extract the values of $H_{c2}$ and $H_{irr}$ using the criterion of 90\% $R_n$ and
10\% $R_n$ respectively, which are shown in Fig. 3(c). It is notable that the in-plane upper critical field $H_{c2}^{ab}$ reveals a very steep increase with cooling near the SC transition. To have a quantitative comparison,
the data are plotted together with that of the bulk sample in Figs. 3(d) and (e). The data of the bulk sample are shifted to lower temperatures to coincide with the $T_c^{onset}$ and $T_c^{zero}$ values of the
ultrathin sample. For the upper critical field, as shown in Fig. 3(d), the behaviors of $H_{c2}^{ab}-T$ is very similar between the ultrathin and bulk samples. Whereas, the $H_{c2}^{c}$ value of the
ultrathin sample rises more gently with cooling near $T_c^{onset}$, as compared with the bulk one, which gives rise to a larger anisotropy $\Gamma = H_{c2}^{ab}/H_{c2}^{c}$ for the ultrathin sample. The detailed values of the slops $d\mu_0H_{c2}^{ab,c}(T)/dT|_{T_c}$, along with
the information about $T_c$, are summarized in Table I. One can see that the value of $d\mu_0H_{c2}^{ab}(T)/dT|_{T_c}$ only decreases a bit when sample becomes ultrathin. In contrast, $d\mu_0H_{c2}^{c}(T)/dT|_{T_c}$
is reduced almost by half. The in-plane irreversible field $H_{irr}^{ab}$ is clearly suppressed in the ultrathin sample, resulting in a more broadening flux liquid region when the field is applied within the $ab$ plane.
Meanwhile, the out-of-plane $H_{irr}^{c}$ reveals no difference with the bulk sample in the low field region below 2 T and is slightly lower in the higher filed region.

\begin{figure}
\includegraphics[width=7.5cm]{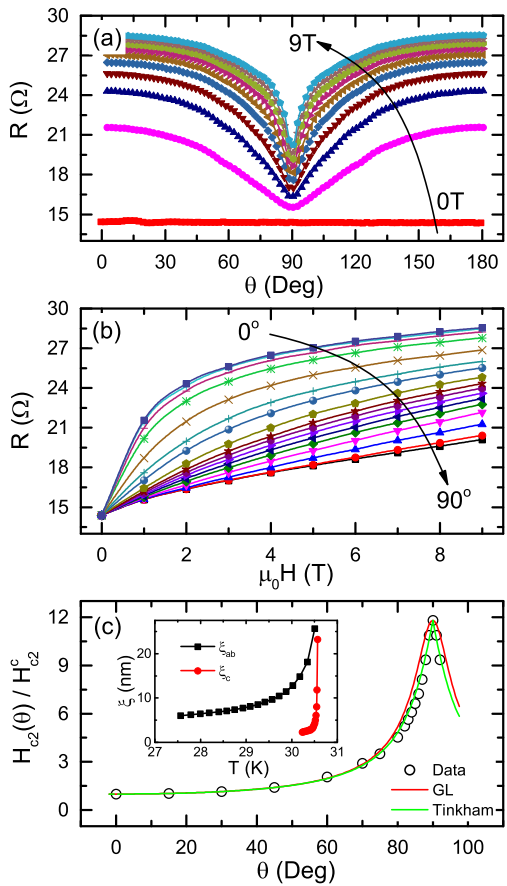}
\caption {(color online) (a) Magnetic-field-angular-dependence of electronic
resistance for the ultrathin sample S-2 at 29.5 K under magnetic fields up to 9 T. The increasing steps of the field is 1 T. (b) Field dependence of the resistance with the
external field rotating from $H$//$c$ ($\theta$ = 0$^o$) to $H$//$ab$ ($\theta$ = 90$^o$) at a fixed temperature $T$ = 29.5 K. The increasing steps of the
angle are 15$^o$, 5$^o$, and 1$^o$ at angle ranges 0 $\sim$ 60$^o$, 70 $\sim$ 80$^o$, and 82 $\sim$ 90$^o$, respectively. (c) Angle dependence of the upper critical field
$H_{c2} (\theta)$ normalized by $H_{c2}^c$. The values of $H_{c2} (\theta)$ are determined using a criterion of 75\%$R_n$. The
theoretical curves based on the GL model and Tinkham model are shown in comparison with the experimental data. The inset shows the temperature dependence of coherence length close to $T_c$.} \label{fig4}
\end{figure}

To further explore the dimensionality effect of the ultrathin K-12442 system, we carried out the field-angle resolved experiments  by measuring the field and angle ($\theta$) dependence of resistance at a fixed
temperature 29.5 K. As shown in Fig. 4(a), the clear anisotropy of the upper critical field results in the variation of resistance with the direction of applied field. From this data, we can obtain the field dependence
of resistance at different angles, which is shown in Fig. 4(b). One can see how the
resistance is enhanced by the field and even saturates gradually with the increase of field in the low angle region. Due to the very large slope of $H_{c2}^{ab}$ with $T$, the increase of resistance is rather limited even
under a field of 9 T. Thus a criterion of 75\%$R_n$ is adopted to define the upper critical fields. It is notable that the detailed criterion does not change the behavior fundamentally (see SI).
Angle dependence of $H_{c2} (\theta)$ normalized by
$H_{c2}^{c}$ is shown in Fig. 4(c), from which one can see the detailed evolution of $H_{c2} (\theta)/H_{c2}^{c}$ versus $\theta$. In order to quantitatively evaluate such an
angle dependent variation, we employ two theoretical models, the 3D GL model~\cite{GL} and 2D Tinkham model~\cite{Tinkham} (see SI), and plot the
curves based on them with the experimental data for comparison. These two models are almost the same in the low-angle region and show
slight differences above 80$^o$, which is usually used to distinguish the 2D superconductivity in ultrathin or interfacial systems~\cite{Ising-3,MoS2-1,2D1,Ma2018}.
It is clear that the Tinkham model can better described the experimental data, indicating the 2D characteristic of the superconductivity in the present ultrathin K-12442 system.

Such a 2D feature was not observed in the bulk sample. In the bulk K-12442 system, angle dependence of $H_{c2} (\theta)$ follows a tendency of the 3D GL model (see SI). The evolution from 3D in the bulk sample to 2D
in the ultrathin sample represents clearly the dimensionality effect in this system. To quantitatively understand this behavior, we estimated the values of coherence length $\xi$ near $T_c$ from the upper critical field (see inset
of Fig. 4(c), details in the estimation see SI). Due to quick increase of $H_{c2}^{ab}$ near $T_c$, we can only obtain the values of out-of-plane $\xi_{c}$ above 30 K. Nevertheless, we still grasp important information that $\xi_{c}$ is several nanometers at around
30 K and increases quickly in higher temperatures, which is in the same order of magnitude with the thickness of the ultrathin sample (2.6$\sim$5 nm). This is the internal reason for the observation of the 2D superconductivity.
The clear dimensionality effect reveals the importance of interlayer coupling in this iron-arsenic system, especially when compared with cuprates superconductors.

\section*{4 Conclusions}
In summary, we successfully obtained the ultrathin samples of the iron-arsenic superconductor KCa$_2$Fe$_4$As$_4$F$_2$ by using an improved mechanical exfoliation method. The dimensionality effect was observed
in terms of the lower superconducting transition temperature, the wider transition width, the broadening flux liquid region, and more importantly, the 2D feature in the upper critical field. Our results reveal the
different features between the iron-arsenic and cuprates superconductors, and provide a suitable platform to investigate the dimensionality effect in unconventional superconductors.

\begin{acknowledgments}
This work is supported by the Youth Innovation Promotion Association of the Chinese
Academy of Sciences (No. 2015187) and the Natural Science Foundation of China
(Nos. 11204338, 11574338).
\end{acknowledgments}

\bibliography{12442-Ultrathin}

\end{document}